\documentclass[a4paper]{article}
\usepackage{geometry}
\usepackage{amsmath,graphicx,color}  
\usepackage{amssymb}
\usepackage{subcaption}
\usepackage{array}
\usepackage{bm}
\usepackage{csquotes}
\usepackage{hyperref}

\newcommand{\eqnref}[1]  {equation~(\ref{#1})}
\newcommand{\figref}[1]  {Fig.~\ref{#1}}

\title{On Kalman-Bucy filters, linear quadratic control and active inference}
\author{Manuel Baltieri$^{1}$, Christopher L. Buckley$^{2}$ \\
\mbox{}\\
$^1$ Laboratory for Neural Computation and Adaptation, \\ RIKEN Centre for Brain Science, Saitama, Japan \\
$^2$ Evolutionary and Adaptive Systems Research Group and Sussex Neuroscience, \\ Department of Informatics, University of Sussex, Brighton, UK \\ \\
manuel.baltieri@riken.jp} 

\begin{document}

\maketitle

\begin{abstract}
Linear Quadratic Gaussian (LQG) control is a framework first introduced in control theory that provides an optimal solution to linear problems of regulation in the presence of uncertainty. This framework combines Kalman-Bucy filters for the estimation of hidden states with Linear Quadratic Regulators for the control of their dynamics. Nowadays, LQG is also a common paradigm in neuroscience, where it is used to characterise different approaches to sensorimotor control based on state estimators, forward and inverse models. According to this paradigm, perception can be seen as a process of Bayesian inference and action as a process of optimal control. Recently, active inference has been introduced as a process theory derived from a variational approximation of Bayesian inference problems that describes, among others, perception and action in terms of (variational and expected) free energy minimisation. Active inference relies on a mathematical formalism similar to LQG, but offers a rather different perspective on problems of sensorimotor control in biological systems based on a process of biased perception. In this note we compare the mathematical treatments of these two frameworks for linear systems, focusing on their respective assumptions and highlighting their commonalities and technical differences.
\end{abstract}

\section{Introduction}
The Linear Quadratic Regulator (LQR) is is a method originally developed in the field of control theory \cite{kalman1960general, kalman1960contributions, anderson1990optimal, stengel1994optimal} with the goal of describing a solution to linear control problems using ``full-state'' feedback methods, i.e., methods relying on a state of the system to regulate (or plant) that is fully available. In more realistic applications however, it is common to assume that the state of a plant and its ensuing dynamics cannot be fully observed, i.e., they are only \emph{partially} observable, or \emph{hidden}. With incomplete information, an \emph{observer} (or estimator) is usually introduced to first infer the hidden dynamics of a system, combining this inference on the (hidden) state of a plant with LQR methods to build an optimal feedback law \cite{aastrom1970introduction, anderson1990optimal, stengel1994optimal}. In the case of linear continuous time problems, the optimal observer has the form of a Kalman-Bucy filter \cite{jazwinski1970stochastic, anderson1972more, stengel1994optimal}. The combination of Kalman-Bucy filters and LQR often goes by the name of Linear Quadratic Gaussian (LQG) controllers \cite{wonham1968separation, aastrom1970introduction, anderson1990optimal, stengel1994optimal}. Despite some of its well-known limitations (e.g., its robustness issues \cite{doyle1978guaranteed}), LQG constitutes nonetheless a standard approach for a large class of linear problems in \emph{stochastic} optimal control \cite{aastrom1970introduction}, due to its analytical tractability and guaranteed optimality under a given set of assumptions. In the last few decades, LQG and some of its extensions have also become a standard paradigm outside of the field of control theory, in particular for models of sensorimotor control in neuroscience \cite{loeb1990understanding, todorov2002optimal, todorov2005stochastic, franklin2011computational}. This reflects recent developments in computational models of perception and behaviour, nowadays mostly described in terms of state estimation, or inference \cite{knill1996perception, rao1999predictive, lee2003hierarchical, knill2004bayesian, Friston2010nature}, and optimal control \cite{kawato1999internal, wolpert2000computational, todorov2002optimal, todorov2004optimality, franklin2011computational} respectively.

In the last few years, active inference has also been introduced in computational and cognitive neuroscience as a theory to describe brain functioning in terms of (approximate Bayesian) inference and control \cite{Friston2010biocyb, Friston2010nature, friston2015active, friston2017active, parr2018discrete}. Similarly to LQG, active inference has also been instrumental in defining models of perceptual and motor processes, from physiological accounts of motor control \cite{adams2013predictions}, all the way to computational descriptions of optimal behaviour claimed to be contrasting LQG \cite{friston2011optimal}. The rapidly evolving nature of this theory has so far encouraged different authors to cover and summarise some of its different technical aspects \cite{bogacz2017tutorial, buckley2017free, biehl2018active, sajid2019active, gershman2019does, da2020active}. At the moment, however, the relationship between this framework and popular LQG-based models in neuroscience remains still unclear, especially considering how the two share a common mathematical background based on methods coming from (approximate) Bayesian inference and stochastic optimal control, but generate rather different predictions. The goal of this note is thus to produce a comparison of the mathematical treatments introduced by active inference and LQG. The discussion will be based on 1) continuous time linear problems, using 2) standard formulations of both frameworks (thus not explicitly addressing more advanced proposals, such as the ``optimal feedback control'' extension of LQG \cite{todorov2002optimal, todorov2005stochastic}, leaving this for future work), to focus on 3) their possible implications for the neural and the cognitive sciences.

\section{Linear Quadratic Gaussian control for linear multivariate continuous time systems}
\label{apdx:KB-LQR-LQG}
In this section we introduce the mathematical underpinnings of LQG control, starting from a general formulation of the Kalman-Bucy filter and LQR for multivariate systems in continuous time, and then combining their results under the separation principle \cite{wonham1968separation} to form LQG controllers. A full derivation, including mathematical proofs and other technical details not relevant to our comparison can be found in standard treatments of Kalman filtering \cite{jazwinski1970stochastic, chen2003bayesian} and LQR/LQG \cite{anderson1990optimal, stengel1994optimal}.

\subsection{Kalman-Bucy filter or Linear Quadratic Estimator (LQE)}
\label{apdx:KalmanFilter}
The Kalman-Bucy filter, also known as Linear Quadratic Estimator (LQE), is an algorithm for the estimation of hidden or latent variables evolving over time, i.e., hidden states with some (linear) dynamics \cite{kalman1960new, sorenson1970least, jazwinski1970stochastic, chen2003bayesian}. Its popularity is mainly due to the fact that
\begin{itemize}
    \item it's the optimal\footnote{In a least-mean square/maximum likelihood sense and, for the conditions met by LQG control and adopted in this note, also minimum variance and unbiased \cite{jazwinski1970stochastic, chen2003bayesian}.} filter for estimation/inference of hidden states in linear systems,
    \item it provides a recursive algorithm well suited for computer simulations and online applications with new data added over time, and
    \item it makes no assumptions on the stationarity of the hidden dynamics a system, particularly relevant for real-world applications and especially control problems.
\end{itemize}
In this treatment we will focus on the continuous formulation of this algorithm, the Kalman-Bucy filter \cite{kalman1961new}, as opposed to its discrete formulation, often simply going by the name of Kalman filter \cite{kalman1960new}. To introduce the Kalman-Bucy filter, we initially define a (linear) continuous dynamical systems in state-space form, representing the simplest plant whose hidden state the filter ought to estimate:
\begin{align}
	d \bm{x} = & A \bm{x} \: dt + d \bm{w} \nonumber \\
	\bm{y} = & C \bm{x} + d\bm{z}
\end{align}
Here the bold characters indicate vectors. The first equation describes linear dynamics for a vector of hidden states $\bm{x} \in \mathbb{R}^n$. In the second equation, $\bm{y} \in \mathbb{R}^m$ is a vector of noisy measurements, or observations, of hidden states $\bm{x}$. Vectors $\bm{w} \in \mathbb{R}^r, \bm{z} \in \mathbb{R}^s$ correspond to Wiener processes for state and observation equations respectively, with $d \bm{w} = \mathcal{N}(0, \Sigma_w), d\bm{z} = \mathcal{N}(0, \Sigma_z)$ thus defined as zero-mean white Gaussian random variables with covariance matrices $\Sigma_w, \Sigma_z$ \cite{jazwinski1970stochastic, chen2003bayesian}. Matrices are represented using capital letters, with $C \in \mathbb{R}^{m \times n}$ as the observation matrix mapping dynamics $\bm{x}$ to observations $\bm{y}$ and $A\in \mathbb{R}^{n \times n}$ as the state transition matrix characterising the time-dependent evolution of $\bm{x}$.

The Kalman-Bucy filter is known in the literature to be the optimal estimator for linear systems with quadratic cost functions and Gaussian white random variables \cite{kalman1960new, jazwinski1970stochastic, chen2003bayesian, aastrom2010feedback}. Under these assumptions, it is also a minimum variance estimator \cite{jazwinski1970stochastic, chen2003bayesian}, i.e., it minimises an objective function represented by the mean square error (MSE), or variance of the error, given by:
\begin{align}
  J = E [(\bm{x} - \hat{\bm{x}})^T (\bm{x} - \hat{\bm{x}})]
  \label{eq:KalmanFilterMSE}
\end{align}
where $\hat{\bm{x}}$ is a vector of the estimates of states $\bm{x}$. Through the Kalman-Bucy algorithm, one can determine estimates $\bm{x}$ via \cite{jazwinski1970stochastic, stengel1994optimal, chen2003bayesian}:
\begin{align}
    \dot{\hat{\bm{x}}} = & A \hat{\bm{x}} + K (\bm{y} - C \hat{\bm{x}}) \nonumber \\
    K = & P C^T (\Sigma_z)^{-1} \nonumber \\
    \dot{P} = & \Sigma_w + A P + P A^T - K (\Sigma_z) K^T
    \label{eq:KalmanBucy}
\end{align}
The vector $\hat{\bm{x}}$ is thus updated using the sum of the current best estimate multiplied by the known transition matrix, $A\hat{\bm{x}}$, and prediction errors (or innovations), $\bm{y} - C \hat{\bm{x}}$. These prediction errors are multiplied by the \textbf{Kalman gain} matrix, $K \in \mathbb{R}^{n \times m}$, expressed in the second equation, and representing the optimal trade-off between previous estimates and information gathered from new observations. To calculate the Kalman gain matrix one then needs to estimate $P \in \mathbb{R}^{n \times n}$, the a-posteriori error covariance matrix, expressing the accuracy of the state estimate in the first equation. The trace of $P$ (i.e., the sum of the components on the main diagonal) gives the sum of the independent components of the covariance matrix equivalent to the MSE in \eqnref{eq:KalmanFilterMSE}, thus describing the accuracy of the state estimation process too. Kalman(-Bucy) filters can also be described in terms of Bayesian inference, with $\hat{\bm{x}}$ and $P$ representing mean and covariance matrix of an estimated (multivariate) Gaussian distribution of hidden states $\bm{x}$ \cite{jazwinski1970stochastic, meinhold1983understanding, chen2003bayesian}.

\subsection{Linear Quadratic Regulator (LQR)}
\label{apdx:LQR}
Linear Quadratic Regulator (LQR) is defined for deterministic linear systems with quadratic cost functions. Under these assumptions, the optimal control law is equivalent to a negative (proportional) feedback mechanism\footnote{In the case of infinite-horizon control problem.} \cite{kalman1960general, kalman1960contributions, anderson1990optimal}. The noiseless plant to regulate is represented by a linear differential equation:
\begin{align}
    d \bm{x} = A \bm{x} \: dt + B \bm{a} \: dt
\end{align}
with $\bm{x} \in \mathbb{R}^n$ as a vector of measured variables to be controlled. In this case we assume that the state of the plant is directly observable (no measurement noise, $d \bm{z}$ in the LQE formulation) so while the differential form $d \bm{x}$ is not strictly necessary (i.e., we could replace $d \bm{x}$ with $\bm{\dot{x}}$ and drop the $dt$ notation), it is maintained for consistency with the Kalman-Bucy filter definition provided previously. $A \in \mathbb{R}^{n \times n}$ is the state transition matrix, as in the case of the Kalman-Bucy filter, while $B \in \mathbb{R}^{n \times p}$ is the control matrix mapping actions $\bm{a} \in \mathbb{R}^p$ to outputs $\bm{x}$. Actions are often represented by $\bm{u}$ in the control theory literature, however here we highlight a difference that will become important later: in the present note, $\bm{u} \in \mathbb{R}^l$ describes inputs from a more general state-space perspective, i.e., \textbf{all} the variables that affect the (hidden) states of a system but are not states themselves, such that
\begin{align}
	\bm{u} \in \mathbb{R}^l = \{\bm{a} \in \mathbb{R}^p, \bm{d} \in \mathbb{R}^q\} \quad \text{with} \;  l = p + q
	\label{eq:inputs}
\end{align}
In this formulation, $\bm{a}$ is used to represent only a part of all possible inputs, i.e., the subset of inputs produced by a regulator that have an effect on the hidden state of a system (in the sense on Kalman's controllability of a plant \cite{kalman1960contributions, stengel1994optimal}), such as the motor actions of an agent that affect its environment. Variables $\bm{d} \in \mathbb{R}^q$ include, on the other hand, disturbances that influence a system but cannot be governed by a controller. Often, the definitions of $\bm{u}$ and $\bm{a}$ are often used interchangeably, assuming either that 
\begin{itemize}
    \item all the modeled inputs are controllable, i.e., there are no other unknown external forces affecting the state of a system, only small and especially brief perturbations that can be considered as part of noise on dynamics ($d \bm{w}$) \cite{loeb1990understanding, he1991feedback}, or that
    \item perturbations are control-dependent, i.e., they are only due to motor actions and can thus not account for other exogenous sources \cite{todorov2002optimal, todorov2005stochastic}.
\end{itemize}

In LQR, the goal is to stabilise (control, compensate or regulate) variables $\bm{x}$ around target values defined by $\bm{v} \in \mathbb{R}^p$\footnote{Here we want to highlight that in general $p \ne n$, meaning that the degree of controllability of a system, or rather the dimension of action vector $\bm{a}$ to reach target values $\bm{v}$, need not necessarily coincide with the number of states in $\bm{x}$.}, controlling $\bm{x}$ through actions $\bm{a}$. Such actions are determined via the optimisation of a function that accumulates costs over time, called cost-to-go or value functional:
\begin{align}
    J = \int_0^\infty \frac{1}{2} (\bm{x} - B \bm{v})^T Q (\bm{x} - B \bm{v}) + \frac{1}{2} \bm{a}^T R \bm{a} \; dt
    \label{eq:costToGo}
\end{align}
which represents the infinite horizon simplification of the problem, i.e., the upper limit of the integral is infinity. The instantaneous version, simply called cost function or cost rate, is defined for LQR as:
\begin{align}
    c(\bm{x}, \bm{a}) = \frac{1}{2} (\bm{x} - B \bm{v})^T Q (\bm{x} - B \bm{v}) + \frac{1}{2} \bm{a}^T R \bm{a}
    \label{eq:costrateLQR}
\end{align}
with $Q \in \mathbb{R}^{n \times n} \ge 0$ and $R \in \mathbb{R}^{p \times p} > 0$ as arbitrary matrices representing the relative balance between the minimisation of the distance from the target and costs for control respectively. In LQR, the optimal action vector $\bm{a}$ is computed using \cite{anderson1990optimal, stengel1994optimal, todorov2006optimal}:
\begin{align}
    \bm{a} = & - L (\bm{x} - B \bm{v}) \nonumber \\
    L = & R^{-1} B^T V \nonumber \\
    - \dot{V} = & Q + A^T V + V A - L^T R^{-1} L
    \label{eq:LQR}
\end{align}
The first equation implements a negative feedback mechanism on $\bm{x}$ using actions $\bm{a}$. In the second equation, $L$ is the \textbf{feedback gain} matrix, determined using the matrix $V$, which is the Hessian of the cost-to-go functional defined in \eqnref{eq:costToGo} \cite{todorov2008general}.

\subsection{Linear Quadratic Gaussian (LQG) control}
\label{apdx:LQG}
One of the limitations of standard LQR controllers lies in the fact that they do not explicitly deal with state/observation uncertainty (or noise), i.e., the original formulation is for deterministic systems only \cite{kalman1960contributions}. On the other hand, in real-world engineering applications as well as in biological systems, it is more common to think of systems with limited access to information from the environment and actions $\bm{a}$ thus applied to a set of hidden states $\bm{x}$ when only noisy measurements $\bm{y}$ are available. In the control theory literature, Linear Quadratic Gaussian (LQG) control \cite{aastrom1970introduction, athans1971role, anderson1990optimal, stengel1994optimal} is thus introduced as a combination of estimation and control for linear systems, building on the previously defined Kalman-Bucy filter and LQR. Under a particular set of assumptions, LQG controllers can then be seen as ``modular'', with estimation and control processes independently designed to form an optimal solution following the \emph{separation principle} of control theory \cite{joseph1961linear, wonham1968separation,georgiou2013separation}. The idea behind the separation theorem is very closely related to the \emph{certainty equivalence} described in econometrics and decision making \cite{simon1956dynamic, theil1957note}, although some works, including \cite{whittle1981risk, bar1974dual, stengel1994optimal}, highlight their differences especially in the context of ``dual effects'' (see Discussion) and risk-sensitive control. In information theory, Shannon \cite{shannon1948mathematical} also introduced a different notion of a separation principle, to explain coding via two (separate) phases of source compression and channel coding \cite{gastpar2003code}. The connections between Shannon's work and the separation principle in control theory have become more clear in recent years, thanks to a growing literature showing how Shannon's definition captures and potentially generalises the results from control theory, see for instance \cite{tatikonda2000control, tanaka2015lqg, fox2016minimum}. Here however, the focus will be on the principle traditionally described in control theory for LQG in continuous systems \cite{joseph1961linear, wonham1968separation}, under the following standard assumptions \cite{wonham1968separation, aastrom1970introduction, bar1974dual, anderson1990optimal, stengel1994optimal}:
\begin{enumerate}
  \item linear process dynamics and observation laws describing the environment and its latent variables,
  \item Gaussian white additive (cf. \cite{todorov2005stochastic}) noise in both process and measurement equations,
  \item known covariance matrices for both process and measurement noise,
  \item quadratic cost function used to measure the performance of a system, and
  \item known inputs, $\bm{u} = \{\bm{a}, \bm{d}\}$.
\end{enumerate}
The last condition is particularly relevant for a comparison with active inference and is due to the assumptions behind the Kalman-Bucy filter, defined as an optimal estimator in an \emph{unbiased} sense only for known inputs \cite{friedland1969treatment}, outputs (i.e., measurements) and parameters. To provide a more formal treatment of the separation principle, we then define a general linear system to be regulated in the presence of noise or uncertainty:
\begin{align}
	d \bm{x} = & A \bm{x} \: dt + B \bm{a} \: dt + d \bm{w} \nonumber \\
	\bm{y} = & C \bm{x} + d\bm{z}
    \label{eq:LQGSSM}
\end{align}
where all the variables and parameters follow the notation previously defined for Kalman-Bucy filters and LQR. In this case, the cost rate is modified to deal with a stochastic system with white additive noise on both dynamics and observations. To do so, the standard formulation of LQR is extended, including stochastic variables so to minimise the \emph{expected} cost-to-go \cite{stengel1994optimal, todorov2006optimal}:
\begin{align}
    J = \mathbb{E} \Bigg[ \int_0^\infty \frac{1}{2} (\bm{x} - B \bm{v})^T Q (\bm{x} - B \bm{v}) + \frac{1}{2} \bm{a}^T R \bm{a} \Bigg]
    \label{eq:costToGoStochastic}
\end{align}
Importantly, one can show that under the assumptions introduced above
\begin{align}
    c(\bm{x}, \bm{a}) = c(\hat{\bm{x}}, \bm{a}) = \frac{1}{2} (\hat{\bm{x}} - B \bm{v})^T Q (\hat{\bm{x}} - B \bm{v}) + \frac{1}{2} \bm{a}^T R \bm{a}
    \label{eq:costrateLQG}
\end{align}
where we replaced states $\bm{x}$ with their estimates $\hat{\bm{x}}$, thus implying that the optimal control can be computed using only the state (point) estimates, i.e., means $\hat{\bm{x}}$. Minimising the expected value of the cost-to-go is thus equivalent to minimising the cost-to-go for the expected state.

\begin{figure}[ht!]
  \centering
  \includegraphics[width=.8\linewidth]{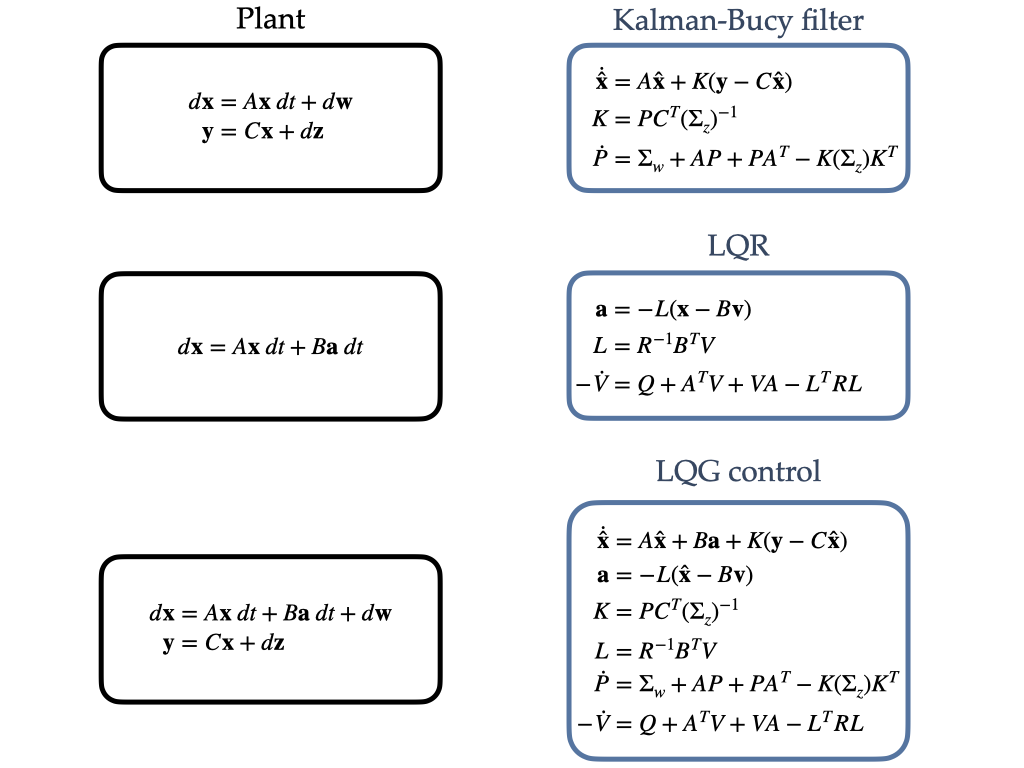}
  \caption{\textbf{Equations for Kalman-Bucy filter, LQR and LQG.} Here we summarise the Kalman-Bucy filter, LQR and LQG algorithms and their respective plants.}
  \label{fig:LQGSummary}
\end{figure}

Optimal estimation and control are then achieved by sequentially combining two separate parts, a Kalman-Bucy filter and a LQR, to produce a regulator optimal in a minimum-variance, unbiased sense \cite{anderson1990optimal, stengel1994optimal}. One thus obtain the following scheme:
\begin{align}
    \label{eq:LQGEstimationControl}
    \dot{\hat{\bm{x}}} = & A \hat{\bm{x}} + B \bm{a} + K (\bm{y} - C \hat{\bm{x}}) \nonumber \\
	\bm{a} = & - L (\hat{\bm{x}} - B \bm{v}) \nonumber \\
    K = & P C^T (\Sigma_z)^{-1} \nonumber \\
    L = & R^{-1} B^T V \nonumber \\
    \dot{P} = & \Sigma_w + A P + P A^T - K (\Sigma_z) K^T \nonumber \\
	- \dot{V} = & Q + A^T V + V A - L^T R L
\end{align}
where the Kalman-Bucy filter provides optimal (unbiased) state-estimates $\hat{\bm{x}}$ of observations $\bm{y}$ that are conditionally dependent only on the vector of motor actions $\bm{a}$ that contributed to the generation of the current observations $\bm{y}$ (i.e., at the time of the estimation), thanks to a Markov assumption due to the presence of \emph{white} noise \cite{chen2003bayesian}. The LQR then uses these point estimates to implement an optimal controller that essentially treats the problem as a deterministic one by replacing the state of a plant with its best estimate, the mean, see also \figref{fig:LQGSummary}. The ability to analytically evaluate the expected cost rate using only expectations $\hat{\bm{x}}$, see \eqnref{eq:costrateLQG}, is thus at the core of LQG architectures. In this way, estimation and control processes interact via the sharing of some information in the form of estimates $\hat{\bm{x}}$ to the controller and a copy of actions $\bm{a}$ to the estimator, but are otherwise seen as fundamentally separable modules to be optimised independently.

\section{Active inference for linear multivariate continuous time systems}
\label{apdx:ActiveInference}
Active inference is a process theory describing brain functioning and behaviour in living systems using ideas from Bayesian inference and stochastic optimal control \cite{Friston2010biocyb, Friston2010nature, friston2015active, buckley2017free, friston2017active}. In this section we begin to establish its relations to the LQG architecture, building an active inference solution to the regulation problem of a linear multivariate system. To do so, we first define a \emph{generative model} \cite{beal2003variational, bishop2006pattern} for a system that prescribes 1) the dynamics of the plant to control (see \eqnref{eq:LQGSSM}) and 2) how these relate to observations. The linear generative model is presented in a state-space form, similar to the the one expressed in \eqnref{eq:LQGSSM}, but is a rather different mathematical object: while \eqnref{eq:LQGSSM} describes the ``real'' plant (i.e., its \emph{generative process} \cite{Friston2010biocyb, buckley2017free, friston2017active}), in a generative model we write a set of equations embodying our assumptions regarding a plant that are used to derive update equations that define a controller. Importantly, the mapping between a generative process and a generative model is not necessarily one-to-one, and different generative models can ensure regulation while representing different sets of assumptions \cite{baltieri2019active, tschantz2020learning}. The generative model to regulate the same linear system previously introduced in \eqnref{eq:LQGSSM} is thus defined as\footnote{It should be noted that Wiener processes in continuous time are often presented using a notation in terms of differentials $dx$, $dt$ and $dW$ where $W$ is a Wiener process or Brownian motion, see \eqnref{eq:LQGSSM}. However, it is also possible to simply use a Langevin form as we do here even in the case of white noise, at least until an appropriate calculus (e.g., Ito or Stratonovich) is introduced and the appropriate interpretation applied \cite{van1981ito}. Notice also that we adopt here a Lagrange notation for derivatives, with Newton's notation introduced later to specify the ensemble dynamics of the recognition dynamics, see \cite{buckley2017free}.}:
\begin{align}
    \bm{x}' = & \hat{A} \bm{x}' + \hat{B} \bm{v} + \bm{w} \nonumber \\
    \bm{y} = & \hat{C} \bm{x} + \bm{z}
    \label{eq:GenerativeModel}
\end{align}
where the hat over matrices $\hat{A}, \hat{B}, \hat{C}$ highlights the important fact that such matrices do not (have to) mirror their counterparts describing the world dynamics in \eqnref{eq:LQGSSM}, i.e., as explained above one can describe several generative models for a given plant. In the simplest case, this may simply imply that matrices $A, B, C$ are unknown and should also be estimated \cite{rao1999optimal, Friston2008c}. More in general however, different matrices in a generative model can implement desired dynamics that substantially differ from the observed ones in order to obtain specific desired behaviours\footnote{within reachability and controllability constraints \cite{stengel1994optimal}} as previously shown in for instance \cite{friston2009reinforcement, Friston2010biocyb, baltieri2017active, baltieri2019pid}. As we will discuss later, this constitutes a point of departure from the traditional assumption behind architectures built on the separation principle, where the variables described in LQG controllers reflect, by construction, the linear dynamics of the observed system, such that 
\begin{align}
    \hat{A} = A, \quad \hat{B} = B, \quad\hat{C} = C
    \label{eq:matricesGenModel}
\end{align}
Furthermore, while in LQG one has to assume full knowledge of motor actions $\bm{a}$ for the unbiasedness of Kalman filters \cite{stengel1994optimal}, in active inference this vector is not explicitly modeled, leading to a case where no copy of motor signals is available or even necessary, relying instead on a process of \emph{biased} estimation \cite{baltieri2019nonmodular} (cf. \cite{friedland1969treatment}). Active inference proposes that a deeper duality of estimation and control exists whereby, in the simplest case, actions are just peripheral open-loop responses to the presence of prediction errors on observations, irrespectively of the causes of sensations: self-generated, $\bm{a}$, or external, $\bm{d}$ \cite{friston2011optimal, friston2012agency, brown2013active, adams2013predictions}. While our focus remains on the continuous time formulation of active inference, it should be noted that in more recent discrete formulations of behaviour under this framework, this account was extended with action cast as a problem of minimising \emph{expected} free energy, inferring (fictitious) control states $\bm{c}$ or rather, time-dependent sequences of control states, or policies $\pi_{\bm{c}}$ \footnote{called $\bm{u}$ and $\pi_{\bm{u}}$ in \cite{friston2015active}.} associated to a set of actions $\bm{a}$ \cite{friston2012agency, friston2015active, tschantz2020learning}. Importantly however, both these proposals rely on theories in neuroscience that suggest a lack of information on self-produced controls (i.e., efference copy \cite{von1950reafferenzprinzip}), proposing an alternative for effective motor control in biological systems \cite{feldman2009new, friston2011optimal, adams2013predictions, feldman2016active, latash2019physics}. The generative model thus does not describe directly the role of actions $\bm{a}$ as seen in \eqnref{eq:LQGSSM}. In their place, the vector $\bm{v}$ encodes instead priors in the form of external or exogenous inputs in a general state-space models sense, (cf. $\bm{u}$ in \eqnref{eq:inputs}), or biases in probabilistic settings. In this light, priors can be used as targets for the regulation problem (see \eqnref{eq:costToGo} and \eqnref{eq:costToGoStochastic}), effectively biasing the estimator to infer desired rather than observed states, with a controller instantiating actions on the world to fulfill the target (state or trajectory) of a controller \cite{baltieri2019pid}. Variables $\bm{z}, \bm{w}$ model noise in or uncertainty about the environment in a Stratonovich sense, i.e., having non-zero autocorrelations (see Discussion).

Following the general formulation of the variational free energy under Gaussian assumptions (Laplace and variational Gaussian approximations) applied to the multivariate case, we define (dropping terms constant during the minimisation process):
\begin{align}
    F \approx - \ln P(\bm{y}, \bm{\vartheta}) \Bigr \rvert_{\bm{\vartheta} = \bm{\hat{\vartheta}}}
    \label{eq:freeEnergyLaplaceMultivariate}
\end{align}
with a full derivation found in, for instance, \cite{buckley2017free, daunizeau2017variational, baltieri2019active}. Here $\bm{\vartheta}$ denotes the set of all the hidden variables of a generative model, such that $\bm{\vartheta} = \{\bm{x}, \bm{v}, \bm{\theta}, \bm{\gamma}\}$, with $\bm{x}, \bm{v}$ defined above and parameters $\bm{\theta}$ including the matrices $\hat{A}, \hat{B}, \hat{C}$ previously introduced, while hyperparameters $\bm{\gamma}$ encode the stochastic properties of the generative model, i.e.,  the covariance matrices $\Sigma_z, \Sigma_w$ and their possible reparameterizations.

With Gaussian assumptions on $\bm{z}, \bm{w}$, the state-space model in \eqnref{eq:GenerativeModel} can then be written down in a probabilistic form, mapping the measurements equation to a likelihood
\begin{align}
    P(\bm{y} | \bm{x}, \bm{v}) = P(\bm{y} | \bm{x})
\end{align}
assuming that observations $\bm{y}$ are conditionally independent on inputs $\bm{v}$. The system's dynamics are also then expressed as a prior \cite{Friston2008c}
\begin{align}
    P(\bm{x}, \bm{v}) = P(\bm{x} | \bm{v}) P(\bm{v}) = P(\bm{x} | \bm{v})
\end{align}
where we assumed that inputs $\bm{v}$ are known, with their prior $P(\bm{v})$ thus reducing to distribution with its mass densely concentrated around its mean (in the limit, a delta function). This assumption simply expresses the fact that in our current formulation of control problems, $\bm{v}$ will encode values assigned by us and representing the target-state or trajectory of the system. Since the probabilities densities are both multivariate Gaussian, they can be written as:
\begin{align}
    & P(\bm{y} | \bm{x}) = \frac{1}{\sqrt{(2\pi)^m |\Sigma_z|}} \exp\Big({-\frac{1}{2}(\bm{y} - \hat{C} \bm{x})^T \Sigma_z^{-1} (\bm{y} - \hat{C} \bm{x})}\Big) \nonumber \\
    & P(\bm{x}, \bm{v}) = \frac{1}{\sqrt{(2\pi)^n |\Sigma_w|}} \exp\Big({-\frac{1}{2}(\bm{x'} - \hat{A} \bm{x'} - \hat{B} \bm{v})^T \Sigma_w^{-1} (\bm{x'} - \hat{A} \bm{x'} - \hat{B} \bm{v})}\Big)
    \label{eq:multivariateGaussian}
\end{align}
where $m, n$ represent the dimensions of vectors $\bm{y}$ and $\bm{x}$ and $|\Sigma_z|, |\Sigma_w|$ are the determinants of the respective covariance matrices. By substituting \eqnref{eq:multivariateGaussian} in \eqnref{eq:freeEnergyLaplaceMultivariate}, the variational free energy for a generic linear multivariate system becomes:
\begin{align}
    F \approx & \frac{1}{2} \bigg[ \Big( \bm{y} - \hat{C} \bm{\hat{x}} \Big)^T \Pi_z \Big( \bm{y} - \hat{C} \bm{\hat{x}} \Big) + \Big (\bm{\hat{x}'} - \hat{A} \bm{\hat{\mu}_x} - \hat{B} \bm{\hat{v}} \Big)^T \Pi_w \Big (\bm{\hat{x}'} - \hat{A} \bm{\hat{x}} - \hat{B} \bm{\hat{v}} \Big) - \ln \big| \Pi_z \Pi_w \big| + (m + n) \ln 2 \pi \bigg]
  \label{eq:freeEnergyMultivariate}
\end{align}
where we replaced $\bm{x}, \bm{v}$ with their expectations $\bm{\hat{x}}, \bm{\hat{v}}$, since under the Laplace and the variational Gaussian assumptions, the free energy in \eqnref{eq:freeEnergyLaplaceMultivariate} must be evaluated at the mode of $P(\bm{y}, \bm{\vartheta})$, equivalent to the mean for Gaussian variables. In the same way, we then introduced precision matrices $\hat{\Pi}_z, \hat{\Pi}_w$ as the inverse of the best estimate of covariance matrices $\hat{\Sigma}_z, \hat{\Sigma}_w$, derived from an application of the Laplace approximation to covariances $\Sigma_z, \Sigma_w$ \cite{Friston2008c}. It is important to highlight that, in general, the covariance matrices (or their inverse, the precision matrices) used in the generative model can in fact be different from the ones used to describe the environment or generative process \cite{baltieri2019pid}. The same approximations can in principle be applied to $\bm{\theta} = \{ \hat{A}, \hat{B}, \hat{C} \}$ \cite{Friston2008c}, but to simplify the treatment (and in line with the idea of comparing active inference and LQG on equal ground), we assumed that these parameters are known quantities, even if different from their respective matrices in the generative process (see \eqnref{eq:matricesGenModel}).

\begin{figure}[ht!]
  \centering
  \includegraphics[width=.8\linewidth]{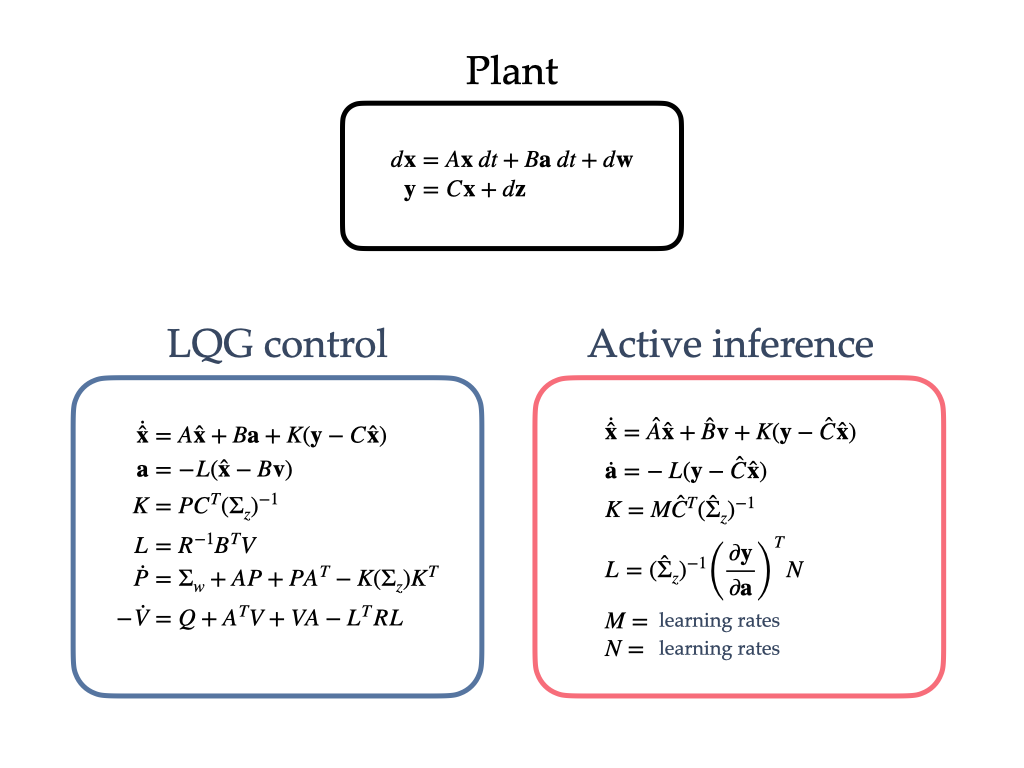}
  \caption{\textbf{LQG and active inference equations.} Different algorithms for the control of continuous time linear partially observable dynamical systems, LQG on the left and active inference on the right.}
  \label{fig:LQGvsAI}
\end{figure}

The recognition dynamics, prescribing estimation and control in a system minimising variational free energy \cite{Friston2008a, Friston2010genfilt, buckley2017free} are implemented in standard active inference formulations as a gradient descent scheme on free energy with respect to $\bm{\hat{x}}$ and $\bm{\hat{x}'}$ for estimation:
\begin{align}
    \bm{\dot{\hat{x}}} & = D \bm{\hat{x}} - \frac{\partial F}{\partial \bm{\hat{x}}} = \bm{\hat{x}'} + \hat{C}^T \Pi_z \Big( \bm{y} - \hat{C} \bm{\hat{x}} \Big) + \hat{A}^T \Pi_w \Big (\bm{\hat{x}'} - \hat{A} \bm{\hat{x}} - \hat{B} \bm{\hat{v}} \Big) \nonumber \\
    \bm{\dot{\hat{x}'}} & = D \bm{\hat{x}'} - \frac{\partial F}{\partial \bm{\hat{x}'}} = - \Pi_w \Big (\bm{\hat{x}'} - \hat{A} \bm{\hat{x}} - \hat{B} \bm{\hat{v}} \Big)
  	\label{eq:perceptionActiveInference}
\end{align}
This system for the update of the means of $\bm{x}$ and $\bm{x'}$ is similar to the standard form of prediction and update steps in a Kalman filter for discrete systems \cite{chen2003bayesian, friston2014cognitive}. The equations include a term $D$, ensuring the convergence of the scheme when higher embedding orders are also optimised, i.e., $\bm{x'}$, see \cite{Friston2008a}. This differential operator is defined so that $D \bm{\hat{x}} = \bm{\hat{x}'} = 0$ and thus gives 0 for all higher orders of derivatives $D \bm{\hat{x}'} = \bm{\hat{x}'', D \bm{\hat{x}''} = \bm{\hat{x}'''} = 0}, \dots$ that are normally assumed to be white noise (see discussion in \cite{Friston2008a}). Its use is equivalent to the presence of a-priori state estimates found in the update equations of traditional treatments of Kalman(-Bucy) filters, and computed in the prediction step. In practice the system in \eqnref{eq:perceptionActiveInference} collapses to a single equation when the update step (second line in \eqnref{eq:perceptionActiveInference}) is assumed to be faster:
\begin{align*}
    \bm{\dot{\hat{x}'}} = 0 \implies \bm{\hat{x}}' = \hat{A} \bm{\hat{x}} + \hat{B} \bm{\hat{v}}
\end{align*}
giving an equation resembling the Kalman-Bucy matrix equation for the estimated mean in \eqnref{eq:KalmanBucy}:
\begin{align*}
    \bm{\dot{\hat{x}}} = \hat{A} \bm{\hat{x}} + \hat{B} \bm{\hat{v}} + \hat{C}^T \Pi_{z} (\bm{y} - \hat{C} \bm{\hat{x}})
\end{align*}
On the other hand, control is defined via a gradient descent on actions $\bm{a}$, assuming the perspective of a system that can only infer that actions have an effect on observations $\bm{y}$ (i.e., no direct knowledge of how $\bm{a}$ affect the latent states $\bm{x}$ or their estimates $\bm{\hat{x}}$):
\begin{align} 
    \bm{\dot{a}} & = - \frac{\partial F}{\partial \bm{a}} = - \frac{\partial F}{\partial \bm{y}} \frac{\partial \bm{y}}{\partial \bm{a}} = - \frac{\partial \bm{y}}{\partial \bm{a}}^T \Pi_{z} \Big( \bm{y} - \hat{C} \bm{\hat{x}} \Big)
  	\label{eq:actionLQGActiveInference}
\end{align}
These equations are finally reported in \figref{fig:LQGvsAI}, with the introduction of learning rates $M$ and $N$ to facilitate a more direct comparison.

\section{A comparison of LQG and active inference formulations}
\label{apdx:LQG-AI}

\subsection{Related work}
In recent years, several reviews \cite{bogacz2017tutorial, buckley2017free, biehl2018active, gershman2019does, da2020active} have tackled different mathematical aspects of active inference, in order to analyse some of its claims and possibly fully uncover its potential as a general theory for the neurosciences. This note however differs in crucial ways from said reviews, offering a technical perspective that should be seen as complementary to previous works. In particular, unlike \cite{bogacz2017tutorial} here we focus on sensorimotor control rather than on pure perceptual accounts of cognitive systems. The current mathematical treatment builds on \cite{buckley2017free} where an explicit derivation of active inference in generalised coordinates of motion for non-Markovian univariate continuous time stochastic processes was introduced, inspired by more general formulations found, for instance, in \cite{Friston2008a, Friston2008c}. Here we extend this formulation to multivariate systems to introduce a comparison with the standard LQG paradigm, inspired by work initially discussed in \cite{baltieri2018modularity, baltieri2019nonmodular, baltieri2019active}. In \cite{biehl2018active, sajid2019active, da2020active} the focus is on discrete time formulations of active inference \cite{friston2015active, friston2017active} based on partially observable Markov decision processes (POMDPs) and thus differ in fundamental ways from the work presented here which, on the other hand, relies on continuous time systems \cite{Friston2008a, Friston2010biocyb, buckley2017free}. Finally, the work presented in \cite{gershman2019does} also focuses on discrete formulations, without including an explicit account of control and behaviour (similar to \cite{bogacz2017tutorial}).

\subsection{Differences}
As seen in previous sections, and summarised in \figref{fig:LQGvsAI}, LQG and active inference share a common mathematical background rooted in Bayesian inference and optimal control. In practice, however, there are a few key differences between these two architectures for linear continuous time systems. Some of them are rather substantial and affect even more general cases, while others end up playing a relatively minor role, especially for the simplified linear systems presented in this note. Here we will discuss, in particular, differences due to:
\begin{itemize}
	\item state-space model formulations
	\item action selection policies
	\item minimisation schemes
	\item cost functionals and general derivations
\end{itemize}

\subsubsection{Different state-space model formulations}
LQG and active inference both rely on state-space model formulations, however while LQG implements a more traditional one, active inference is said to introduce a definition of models in \emph{generalised coordinates of motion}, i.e., higher embedding orders for continuous time models \cite{Friston2008a, Friston2008c}. These models are equivalent to a continuous time version of standard treatments of coloured (Gaussian) noise, i.e., non-Markovian (Markovian of order \emph{N}, or semi-Markovian) processes, in terms of augmented state-space models seen for discrete systems \cite{jazwinski1970stochastic, Friston2008c}. When the noise is not white, the typical strategy involves increasing the dimensionality of the state-space, treating coloured noise as an autoregressive model of independent white noise components that can be explicitly separated by introducing higher embedding dimensions that maintain an overall Markov property \cite{jazwinski1970stochastic}. In generalised state-space models for continuous systems, random variables are treated as analytical noise with non-zero autocorrelation, replacing the (Markovian) Wiener processes typical of Ito's formulations of stochastic variables\footnote{In Ito calculus autocorrelations are strictly equal to zero \cite{jazwinski1970stochastic}.} and adopted in most standard formulations of Kalman-Bucy filters and LQG control. Wiener processes are usually considered a good approximation of coloured noise only when the time scale of an estimator is much slower than the true time scale of the dynamics of an environment/plant. In this light, non-Markovian processes should simply be seen as a natural generalisation of treatments based on Wiener noise, whose derivatives are mathematically not well defined in the continuous limit, thus making it less suited for models whose characteristic time scale more closely matches the real dynamics. This is often the case for systems studied in various areas of engineering \cite{stratonovich1967topics, jazwinski1970stochastic}, physics \cite{fox1987stochastic, van1992stochastic, luczka2005non, gardiner2009stochastic}, ecology \cite{halley1996ecology, vasseur2004color} and neuroscience \cite{Friston2008c, valdes2011effective, li2011generalised}, among others. Generalised state-space models use a Stratonovich' formulation that allows for differentiable (analytical) noise \cite{belyaev1959analytic, stratonovich1967topics}, often thought to be a better model for coloured noise \cite{wong1965convergence, stratonovich1967topics, van1981ito, Friston2008c}. This generalisation is claimed to allow for the treatment of non-Markovian continuous random variables in a more principled way where derivatives of stochastic processes are defined as in traditional (non-stochastic) calculus and included as extra dimensions of the state-space in terms of higher embedding orders (cf. the role of delayed coordinates in embedding theorems \cite{packard1980geometry, takens1981detecting}). These ``extra'' embedding orders represent a set of coordinates expressed in terms of ``position'', ``velocity'', ``acceleration'', etc., for each variable and correspond to a linearised (for convenience \cite{Friston2008c, buckley2017free}) path generated by a Taylor expansion in time of each trajectory.

The differences between generalised and standard state-space models are however less noticeable for the linear case treated in this note, often used as a starting point for models of sensorimotor control in computational and cognitive neuroscience. This is due both to 1) the treatment in terms of white noise adopted in \eqnref{eq:LQGSSM}, making higher embedding orders redundant when the plant's model is known, and to 2) the white noise being \emph{additive}. Under these assumptions, while the theoretical interpretations of Ito and the Stratonovich formulations still differ \cite{van1981ito}, they lead to the same results \cite{jazwinski1970stochastic}. For nonlinear problems the discrepancies between these two approaches become more obvious, at the cost however, of analytical tractability since stronger assumptions are required for the treatment of more general cases. For instance, in LQG architectures, the separation principle heavily relies on the linearity of observations and dynamics, and only weaker versions of this principle are at the moment defined for some special classes of nonlinear systems, e.g., \cite{atassi1999separation}. In active inference, the effects of the local linearity assumption on higher embedding order remains largely unexplored especially for non-smooth processes. More in general, however, while it is sometimes claimed that active inference has a clear edge over traditional approaches Markovian noisem, such as LQG \cite{Friston2008a, Friston2010genfilt}, one can show that non-Markovian systems can be formulated as Markovian ones using an augmented state-space \cite{jazwinski1970stochastic, Friston2008c}, in the same way higher embedding orders increase the number of state equations in active inference \cite{Friston2008c}. Due to this, some comparisons between active inference algorithms and existing methods in the literature may appear at the moment slightly confusing. For instance, in \cite{Friston2008a} we see an analysis comparing (extended) Kalman-Bucy filters and Dynamic Expectation Maximisation (DEM, one of the standard algorithmic implementations of estimation algorithms in active inference). In presence of white noise, the two methods achieve very similar performances, while for non-white noise, DEM is claimed to display better results \cite{Friston2008a}. At the same time, while the number the state-space dimension for DEM was effectively increased via the use of generalised coordinates of motion, the dimension of the state-space of the (extended) Kalman-Bucy filter remained unchanged, thus suggesting that differences are mainly due to the treatments (or lack of) of coloured noise, with the (extended) Kalman-Bucy filter capable of replicating the same results with an adequate state augmentation. 

\subsubsection{Different update equations for action}
As shown in \figref{fig:LQGvsAI}, the update equations for action in LQG and active inference present a few crucial differences. Importantly, actions in LQG are implemented using an algebraic equation equivalent to standard negative \emph{proportional} feedback mechanisms, while in active inference the same update relies, effectively, on an implementation based on \emph{biased} perception and akin to \emph{integral} control \cite{baltieri2019pid}. This is mainly due to the fact that while LQG requires full knowledge of inputs $\bm{a}$ to calculate the best estimates $\bm{\hat{x}}$, active inference relies on an implicit mechanism to handle the lack of such inputs: integral control. In particular, the target value for the control problem, encoded by the vector $\bm{v}$, is introduced as a prior in the form of a constant ``disturbance'', or bias, in the generative model in \eqnref{eq:GenerativeModel}. This unaccounted disturbance generates a biased estimate (cf. \cite{friedland1969treatment}) that encodes the desired state/trajectory of a system, reached via the use of an integral controller that acts to ``regulate'' such bias \cite{johnson1968optimal, johnson1970further}. Integral control can in fact be seen as an implicit mechanism to estimate linear (i.e., constant) perturbations/inputs, inferred by accumulating evidence based on steady-state errors generated by the presence of the bias vector term $\bm{v}$\footnote{An explicit version of this method corresponds to introducing extra variables in a state-space model for the inference of (linear) inputs as shown, for instance, in \cite{johnson1975observers}.}.

A second key point is that actions built in LQG schemes use prediction errors on hidden states, while active inference updates are based on errors on observed states. In LQG, this is due to an assumption on the invertibility of matrix $C$, mapping estimated states $\hat{x}$ to observations $\bm{y}$, see for instance eqn. 4.1-1 to 4.1-8 in \cite{anderson1990optimal}. Thanks to this assumption, the desired/target trajectory can be expressed directly into the frame of reference of hidden rather than observed states, adapting (thanks to the separation principle) a problem of output-feedback control into one of state-feedback. While this assumption holds for the linear systems defined in LQG under the separation principle, it is not trivially generalised to nonlinear cases where processes of perception and control have strong co-dependences \cite{bar1974dual}. On the other hand, active inference implementations can in principle easily provide approximate solutions to increasingly complicated problems of (sensorimotor) control for a reasonable choice of the peripheral (open-loop) ``inverse'' model, $\frac{\partial \bm{y}}{\partial \bm{a}}^T$ \cite{baltieri2019pid}.

Furthermore, it should also be noted that the update equation for action in active inference replaces the arbitrary positive definite weight matrix $R$, standard in LQG control, with the expected precision matrix of observation noise $\hat{\Pi}_z( = \hat{\Sigma}_z^{-1})$. This move can be seen in light of the established duality of deterministic control and stochastic filtering/estimation problems first formalised by Kalman \cite{kalman1960new, kalman1960general, kalman1961new}. Unlike standard uses of this duality theorem, normally simply highlighting matching terms in the solutions of these two problems in the linear case under a coordinate transformation (including time reversal), active inference assumes that the matrix $R$ in fact just -- \emph{is} -- the expected precision matrix of observation noise $\hat{\Pi}_z$. The implications of this idea include, among others, an intrinsic ``dual effect'' of actions within the active inference framework due to the lack the conditions for the separation principle, cf. LQG \cite{bar1974dual}. This dual effect can be seen as an instantiation of the fundamental exploration-exploitation dilemma \cite{feldbaum1965dual}, portraying the time-constrained trade-off between inferring/learning about the unknown structure of a system and regulating its state to a desired target. The active inference formulation for linear systems constitutes thus an intermediate stage between LQG models where dual effects are not present \cite{bar1974dual}, and more general dualities of estimation and control for nonlinear problems found for instance in \cite{mitter2003variational, todorov2008general}. These generalised dualities are related to the formulation of the information filter, a linear estimator propagating the error precision, rather than covariance, matrix. The information filter provides a formulation that simplifies the duality relations by rearranging matching terms, creating new and different dualities, and in particular by replacing a difficult generalisation of the matrix transpose operator with a second time reversal (for details see \cite{todorov2008general}). The same method based on an exponential transformation was likewise applied in path integral control \cite{kappen2005path, kappen2005linear}. In this scheme, one can also show how $R \propto \Sigma_w^{-1}$ (rather than $R \propto \Pi_z$ as in the Kalman duality expressed by active inference in the linear case) for a class of tractable (i.e., linear in $\bm{a}$) fully-observable nonlinear problems where the conditions for the standard separation principle are not met (i.e., LQG emerges only as a special case \cite{kappen2005path}).

\subsubsection{Different minimisation schemes}
The updates proposed with active inference are apparently consistent with Kalman-Bucy filters, although an important difference is clear: the Kalman gain, $K$, and feedback gain, $L$ are not explicitly computed in active inference. The Kalman gain, $K$, and feedback gain, $L$, matrices prescribe the optimal (i.e., minimum variance) update speed of estimates of hidden states and controls, while balancing prior information and new observations. Both $K$ and $L$ require solving Riccati equations involving knowledge of the covariance matrices of dynamics and observation noise in the first case, and weights representing costs for estimation and control in the cost-to-go function for the latter. In current formulations of active inference, the matrices $K$ and $L$ are not as clearly directly defined as they are in LQG architectures. If we consider only the Kalman gain $K$, the main reason is that active inference relies on a series of simplifying assumptions perhaps not best suited to computation, \emph{online}, Kalman and feedback gains. In particular, while the Laplace \cite{beal2003variational} and variational Gaussian approximations \cite{opper2009variational} greatly simplify the variational Bayesian treatment of inference and control problems (and are, together, exact for linear, Gaussian plants), they are usually accompanied by further approximations. These include, for example, the post-hoc optimisation of the covariance matrix of an approximate density under the variational Gaussian approximation, i.e., performed after a certain number of observations \cite{Friston2008a} (although see \cite{Friston2010genfilt} for an approximate online version). In general however, no equations of the Riccati type currently appear in standard formulations of variational free energy minimisation in active inference. Under further assumptions, for example by considering fading-memory Kalman filters, it is however possible to draw a more direct comparison that will be developed in future work. Kalman filters with fading memory can, in fact, be shown to be equivalent to a natural gradient descent on statistical manifolds (\cite{amari1998natural}) in the univariate case, see for instance \cite{ollivier2018online, ollivier2019extended}. Following then the known correspondence (\cite{martens2014new}) between natural gradient and Gauss-Newton type methods used in active inference \cite{Friston2008c} for exponential family models, one can consider cases where active inference emerges as a special (i.e., fading-memory) case of (extended) Kalman-filters, and then discuss a similar approach for the dual equations involving action and the feedback matrix $L$. It should be highlighted that, unlike repeatedly stated in the literature, we believe it is all but clear how the minimisation scheme used by active inference ought to generalise algorithms such as (extended) Kalman-filters beyond, perhaps, the use of higher embedding orders (however see above). At the moment, most practical applications simply replace explicit $K$ and $L$ with approximations based on the use of learning rates, e.g. \cite{baltieri2017active}, or on local linearisation methods \cite{ozaki1992bridge} with varying integration time-steps \cite{Friston2008a}.

\subsubsection{Different cost functionals and derivations}
A further point of departure between these models can be found in the cost functional minimised by the two methods: a value, or cost-to-go, functional for LQG, \eqnref{eq:costToGoStochastic}, and a variational free energy functional for active inference, \eqnref{eq:freeEnergyMultivariate}. The most striking difference is that the former includes a time integral of costs, while the second one doesn't. This difference is crucial in nonlinear cases, or rather in cases where the separation principle does not hold: the minimisation of variational free energy can in fact be seen as giving a ``time-independent'' equilibrium strategy, generating a fixed control strategy independent of future observations \cite{kappen2012optimal} or equivalently, following \cite{bar1974dual}, one can see the minimisation of variational free energy as implementing a \emph{feedback rather than closed-loop} policy that assumes fixed dynamics and costs in the limit of an infinite time horizon \cite{kappen2012optimal}. At the same time, LQG methods based on the separation principle rely on the same (future) time-independence assumption \cite{bar1974dual}, thus implying that the two functionals are not fundamentally different at this level. A more appropriate comparison may be drawn on generalisations of both approaches, path integral control/KL control on one side and the minimisation of \emph{expected} free energy on the other, where active inference is claimed to generalise other control approaches \cite{friston2017active, da2020active}, however this comparison lies outside the scope of this note, focusing instead on linear continuous time and continuous state-space models.

Finally, the free energy functional \eqnref{eq:freeEnergyLaplaceMultivariate} appears to include extra terms when compared to the value function of LQG \eqnref{eq:costToGoStochastic}. The presence of multiple prediction errors in the variational free energy formulation is however just an expression of the probabilistic (i.e., Bayesian) derivation of inference methods \cite{jazwinski1970stochastic}, now generalised to control problems. The different predictions errors in the variational free energy directly map to likelihood and prior distributions as obtained from the joint density of observed and hidden states used to define free energy, see \cite{buckley2017free}. On the other hand, LQG derivations are often based on least-square methods \cite{jazwinski1970stochastic, stengel1994optimal} with no a-priori interpretation of uncertainty, normally included only post-hoc.

\footnotesize
\bibliographystyle{plain}
\bibliography{AllEntries} 

\end{document}